\documentclass[prl,aps,twocolumn,showpacs]{revtex4}

\usepackage{graphicx}
\usepackage{dcolumn}
\usepackage{bm}

\begin{document}

\preprint{APS/123-QED}

\title{Network Structures from Selection Principles}

\author{Vittoria Colizza$^1$, Jayanth R. Banavar$^2$, Amos Maritan$^{1,3}$ and Andrea Rinaldo$^4$}

\affiliation{$^1$ International School for Advanced Studies and INFM, via
Beirut 2-4, 34014 Trieste, Italy \\ $^2$  Department of Physics, The Pennsylvania State University, 104 Davey Laboratory
University Park, PA  16802 USA \\$^3$ The Abdus Salam International
Center for Theoretical Physics, 34014 Trieste, Italy \\ $^4$ Centro Internazionale di Idrologia "Dino Tonini" and Dipartimento IMAGE, Universit\`a di Padova, I-35131 Padova, Italy}

\date{\today}

\begin{abstract}
{\bf We present an analysis of the topologies of a class of
networks which are optimal  in terms of the requirements of having 
as short a route as possible between any two nodes while yet
keeping the congestion in the network as low as possible.
Strikingly, we find a variety of distinct
topologies and novel phase transitions between them on varying the number
of links per node.  Our results suggest that the emergence of the
topologies observed in nature may arise both from growth mechanisms and
the interplay of dynamical mechanisms with a selection process.}

 \end{abstract}

\maketitle

 There have been many exciting recent developments \cite{developments,barab1,barab2} in understanding the
topologies of many natural and artificial 
networks. The analysis of network topology is carried out using classic concepts such as 
clustering \cite{bollobas}, the distribution of the number of links from each node 
(called the degree) \cite{unknown,barab1,barab2} and its small world character \cite{strogatz,newman}. 
Strikingly, many of the observed topologies are quite distinct from those expected for 
generic random networks \cite{erdos,bollobas}. There has been important
progress \cite{barab1,barab2,others,unknown,mendes1} in rationalizing the existence of 
non-universal scale-free networks (the degree distribution exhibits a power law behavior 
over a finite range with a non-universal exponent) by dynamical models 
entailing the growth by node and edge addition (with possible preferential attachment), 
rewiring \cite{barab1} and edge removal \cite{mendes1}.

Our focus here is the proposal and analysis of a class of models in which
the key selection criterion for network topology is optimality.
Channel networks formed in river basins have  been shown to
attain, in the steady state of their dissipative
dynamics epitomized by the general
landscape evolution equation \cite{banavar}, a minimum of total
energy dissipation \cite{rodriguez}. Strikingly, a variety
of robust scaling features emerge that closely resemble those
observed for natural landforms \cite{rodriguez}, and universality
classes exist depending, for example, on the terrain
heterogeneities \cite{river}. Because of the nature of the 
functional to be miminized, all trees,  i.e. networks with no loops,
 are local optima and thus
prevail over networks which are not competitive from an
evolutionary viewpoint \cite{rodriguez,river,banavar}.
 Optimization has been introduced as a possible explanation of the 
degree distribution observed in the Internet topology \cite{fabrikant} 
or to investigate the origin of small-world networks \cite{mathias},
taking into account the physical distance, i.e. Euclidean
distance, between the nodes of a spatial
network.
Scale-free networks arising from optimal design have  been previously
studied \cite{sole}. It has been shown that the minimization of a
linear combination of 
average degree and average distance (the distance between two
nodes is defined as the minimum number of edges traversed
 to join them) can lead to the emergence of a truncated power-law 
in the degree distribution.

Our goal is to understand the topology of networks 
which minimize a physically motivated cost function. Strikingly, 
we find a variety of distinct topologies and novel phase transitions 
between them on varying the number of links per node. 

Suppose that some type of information has to be communicated between pairs
of nodes of the network  \cite{strogatz}. It is plausible that 
besides the average distance between any two nodes, the type of nodes
encountered along the path(s) joining them may also matter in the
optimization of the dynamics of communication taking place in the system. For
example, selective pressure may operate so as to choose certain
nodes because of their high connectedness - or else to avoid them for
the same reason. Associated with the type of node, is a local
feature that depends only on its degree, namely, the number of
edges rooted in the node. On a global scale, we will distinguish
among structures that rewire local features at random  selecting
the changes if the new structure  provides a selective
 advantage. It is well known that in many such optimization 
problems, the key factor that matters is the shape of the cost function
\cite{rodriguez,river}.  The concavity or convexity of the cost function
can be embodied by a power law form with scaling exponent $\alpha$ 
less than or greater than 1 respectively:
\begin{equation} \label{H}
H_{\alpha} \ = \ \sum_{i<j}d_{ij}(\alpha),
\end{equation}
where $i$ and $j$ are pairs of nodes of the network, and
\begin{equation} \label{D}
d_{ij}(\alpha) \ = \ \min_P \sum_{p \in P: i \to j}\,k_p^{\alpha}\,.
\end{equation}
Here $P$ is any path connecting site $i$ to site $j$ of the system, $p$
is any node belonging to such a path and $k_p$ is the degree or
connectivity of node $p$. The weighted distance $d_{ij}(\alpha)$ is a global
quantity associated with the pair $i,j$ and is the minimum of the sum of
degrees $k_p^{\alpha}$ (a local property), evaluated along
the path $P$ from $i$ to $j$, over all the paths
connecting $i$ and $j$. Note that in the special case of
loopless tree-like structures, such a path is unique and
$d_{ij}=\sum_{p \in P: i \to j}\,k_p^{\alpha}$.  In the limiting
case $\alpha \to 0$, Eq.(\ref{D}) becomes the standard definition
of distance on a network \cite{newman}.  The new definition of weighted 
graph distance introduced in
Eq.(\ref{D}) captures the conflict between two
competitive trends: the avoidance of long paths and the desire
to skip heavy traffic. 

The networks minimizing the cost eq.(\ref{H}) are searched for among 
the ensemble containing a 
fixed number of nodes $n$, as well as the number of links (edges) $l$.
The resulting networks are analyzed in terms of the degree distribution
$P(k)$, i.e. the fraction of nodes with degree $k$, the average
distance between pairs of nodes and the  average clustering 
coefficient $C=n^{-1} \sum_i C_i$, where $C_i$
 is a measure of how interconnected the neighbors of a given node 
are \cite{newman}:
\begin{equation} \label{C}
C_i=\frac{l_i}{k_i(k_i-1)/2} \,\,,
\end{equation}
 $l_i$ is the number of links between the neighbours of node $i$ and
$k_i(k_i-1)/2$ is the total number of possible pairs that can be formed
among them.

The optimization method used in the numerical simulations is a
Metropolis scheme at zero temperature. The goal is to obtain the
statistics of all local minima which are
accessible topologies associated with the chosen dynamics \cite{banavar}.

We have studied several values of $\alpha$ and $r = l/n$ with $n = 35-200$.
The protocol of the simulation is as follows:

i) generation of a random initial configuration with fixed $n$ and $l$;

ii) random rewiring: Specifically, a link connecting the
sites $i$ and $j$ is randomly chosen and substituted with a link
from $i$ to a site $k$, not already connected to $i$, extracted
with uniform probability among the sites of the system. This ensures that
the number of links $l$ as well as the size of the system $n$
remains constant during the minimization;

iii) connectedness control: If the graph is not connected after rewiring,
step (ii) is repeated;

iv) energetic control. The new value of $H_{\alpha}(t+1)$ is
calculated. The new configuration is accepted only if it is
energetically favorable, i.e. only if $H_{\alpha}(t+1) <
H_{\alpha}(t)$; otherwise the change is rejected and we return to
step (ii).

Note that the zero-temperature setting ensures feasible
optimality of the emerging network structure \cite{river}, a
feature that is relevant for dynamical accessibility
of complex optimal structures. The minimization algorithm stops
after $F$ consecutive failed changes on the network; we have chosen
$F=n(n-1)$, so that, on average, each pair of vertices is allowed
to change its state  twice. For each case we
performed  $200$ independent simulations, starting with different random
initial configurations and varying the size $n$ of the system:
$n=35,50,70,100,140,200$. For each size, the different values of
the ratio $r$ investigated are:
$r=1.05,1.1,1.2,1.3,2.0,2.3,3.0$.

\begin{figure}[!ht]
\begin{center}
    \includegraphics[width=5.5cm,height=4cm]{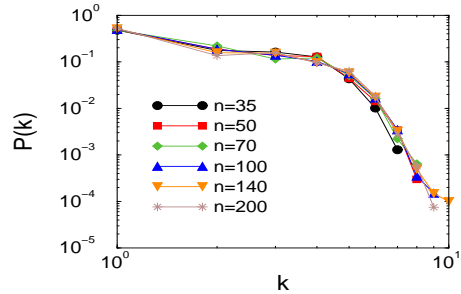}
\end{center}
\vspace{-.7cm} 
\caption{Degree distribution, averaged over 200
realizations, for several system sizes ($n=35,50,70,100,140$) for $\alpha=0.7$
and $r=1.05$. The system displays a range of degrees.} 
\label{fig:Pk_a0.7_1.05__a0.7_n70}
\end{figure}

\begin{figure}[!ht]
\begin{center}
    \includegraphics[width=5.5cm,height=4cm]{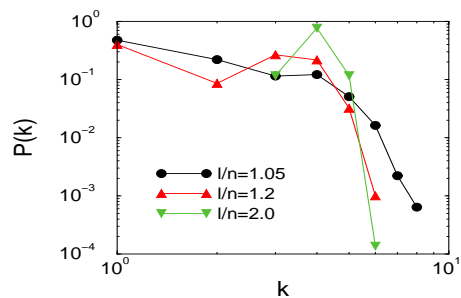}
\end{center}
\vspace{-.7cm} \caption{Crossover between the two distinct
behaviors:   the heterogeneous regime which exhibits a range of 
degrees and the homogeneous  one  characterized by a peaked 
distribution.
Data are averaged over 200  realizations for
$\alpha=0.7$, $n=70$ and for several values of $r=l/n$.}
\label{fig:Pk_n70_a0.7}
\end{figure}

On varying $r$, we observe two distinct behaviors. The first 
occurs for values of $r \sim 1$: the system displays
an apparent scale-free behavior in $P(k)$ 
for several values of $\alpha$ (see Figure \ref{fig:Pk_a0.7_1.05__a0.7_n70}, 
for $\alpha=0.7$). However, the behavior does not seem 
to be a genuine power law because
the sharp cut-off  does not display the expected 
dependence on the system size $n$. Unfortunately, the computational cost, 
which grows exponentially with the number of nodes, does not permit us to 
quantify the weak dependence of the  cut-off on $n$. 
As $\alpha$ increases, this apparent scale-free
region shrinks around the
value $r=1$ and is vanishingly small for $\alpha>1$.
The second  behavior is obtained for larger values of the ratio
$r$ -- the degree distribution obtained is strongly peaked around the
average value of $k$, $<k>$ (Figure (\ref{fig:Pk_n70_a0.7})). 

\vspace{-.4cm}

\begin{figure}[!ht]
\begin{center}
    \includegraphics[width=3.3cm]{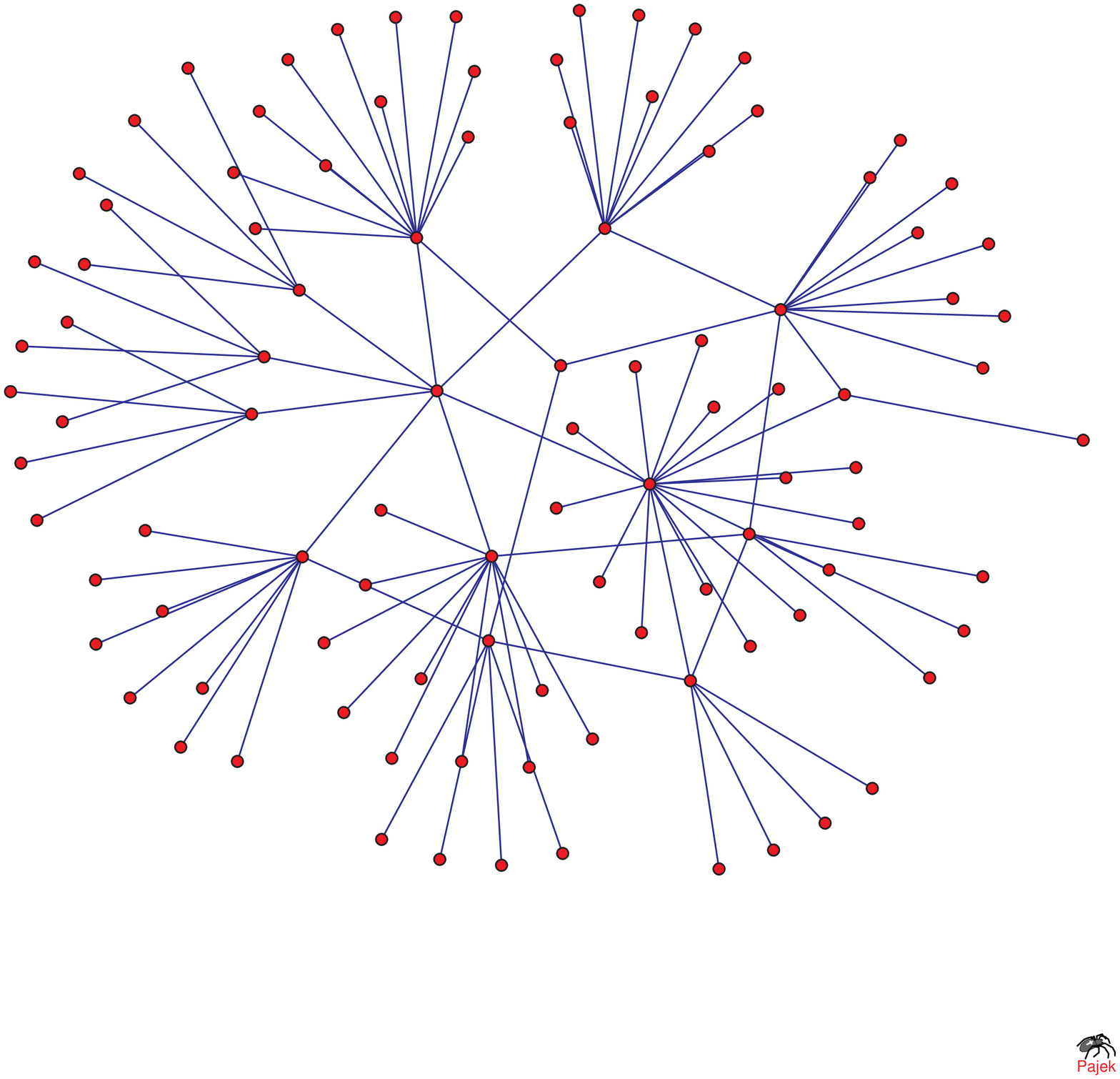}
    \hspace{.5cm}
    \includegraphics[width=3.3cm]{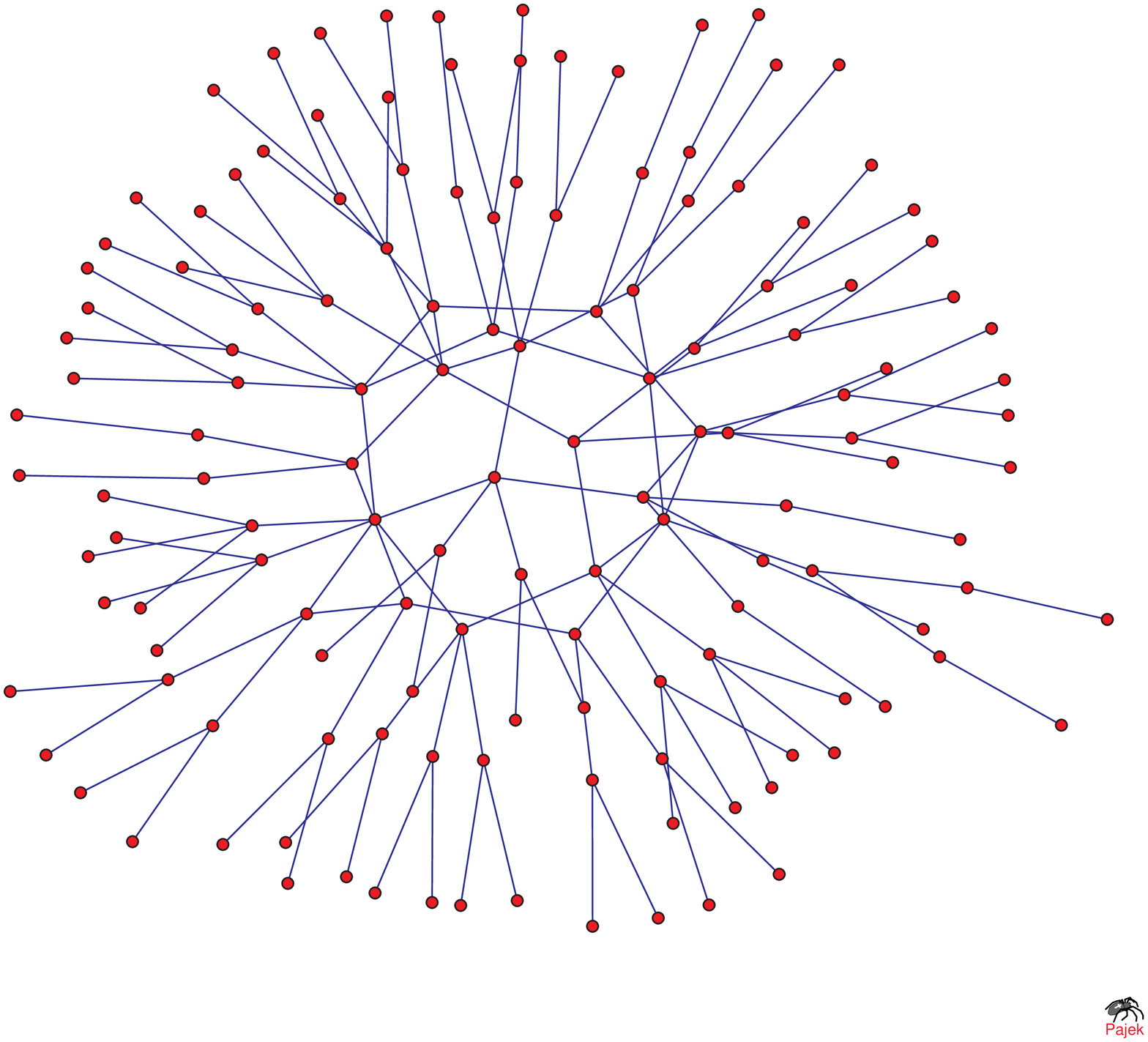}

    \vspace{-.4cm}

    \includegraphics[width=3.3cm]{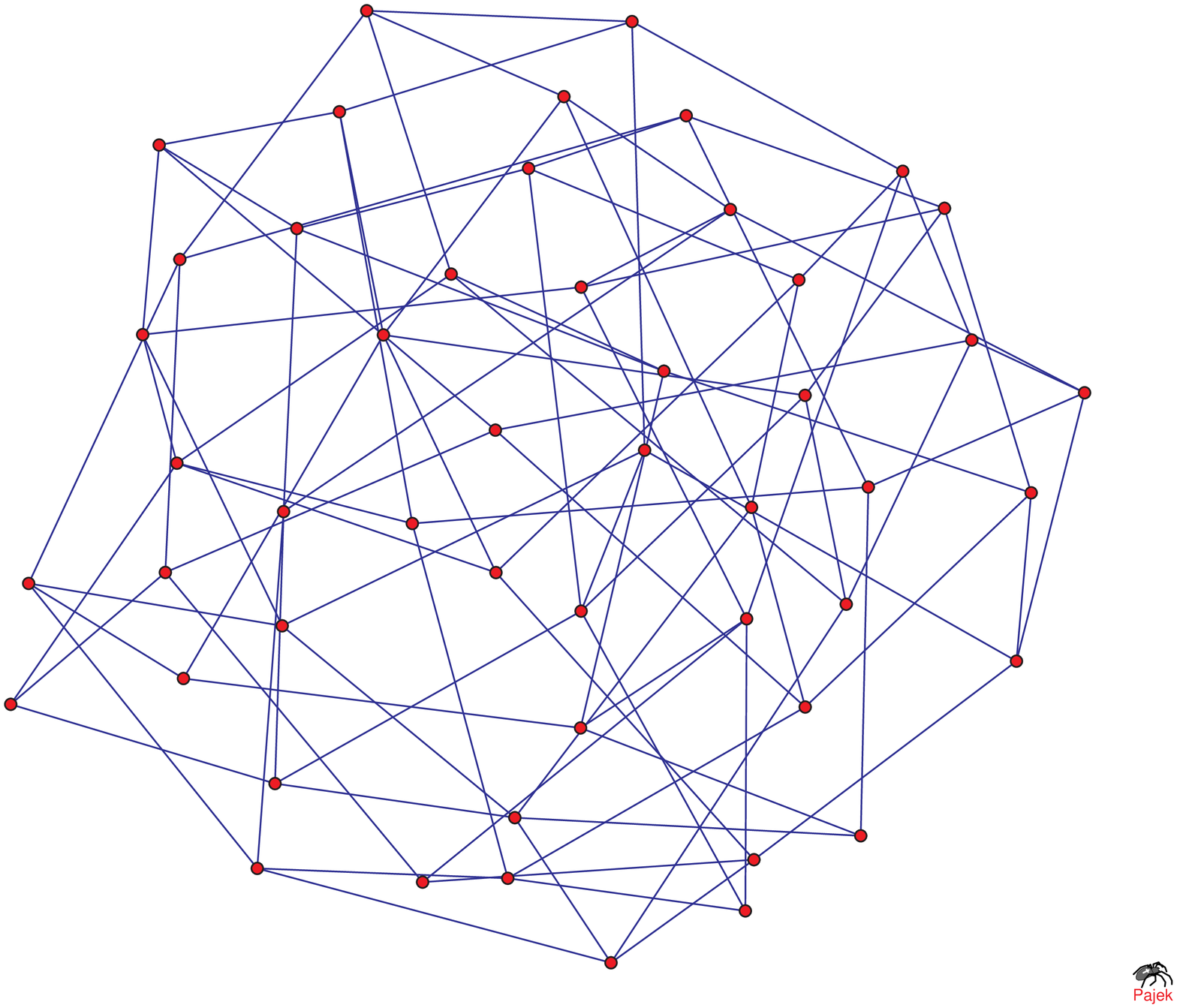}
    \hspace{.5cm}
    \includegraphics[width=3.3cm]{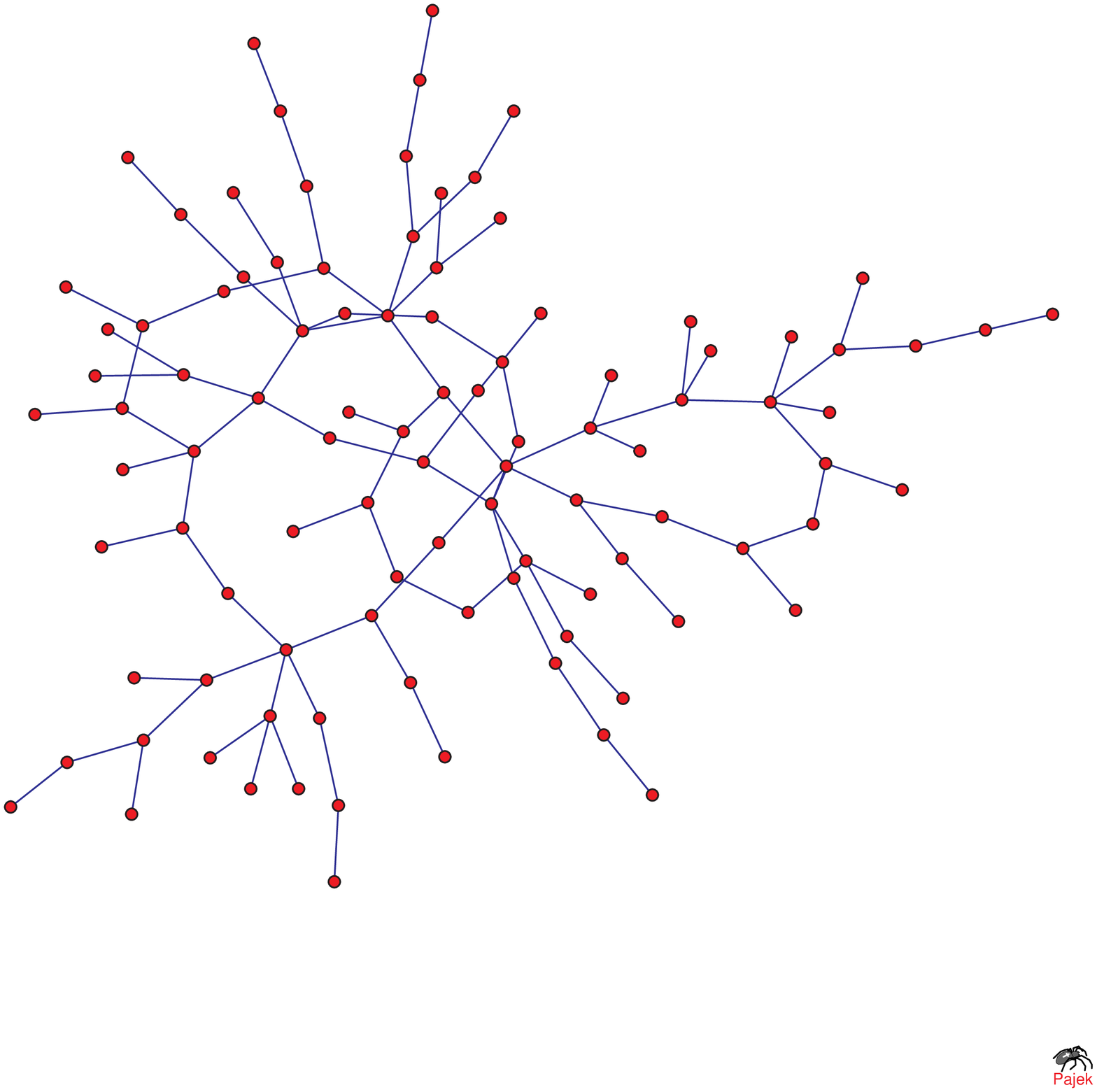}
\end{center}
\vspace{-.7cm}
\caption{Graph representation of four typical networks
with: Top Left:
$\alpha=0.4,\,r=1.05,\,n=100$; Top Right:
$\alpha=0.7,\,r=1.05,\,n=140$; Bottom Left:
$\alpha=0.5,\,r=2.0,\,n=50$; Bottom Right:
$\alpha=2.0,\,r=1.05,\,n=100$. The graphs have been produced with
the Pajek software.} \label{fig:topologies}
\end{figure}

A sample of network topologies 
are illustrated in Figure (\ref{fig:topologies}),
for different values of $\alpha$ and $r$.

On increasing the value of the ratio $r$, one moves from
networks characterized by the presence of some highly
connected nodes together with many peripheral sites (Top Left
and Right) to networks in which almost every node has the same
degree $k=<k>$ (Bottom  Left and Right). In addition,
a sharp transition is observed in terms of
the average clustering coefficient $C=<C_i>$, as defined in eq.(\ref{C}).

\begin{figure}[!ht]
\begin{center}
    \includegraphics[width=5.5cm,height=4cm]{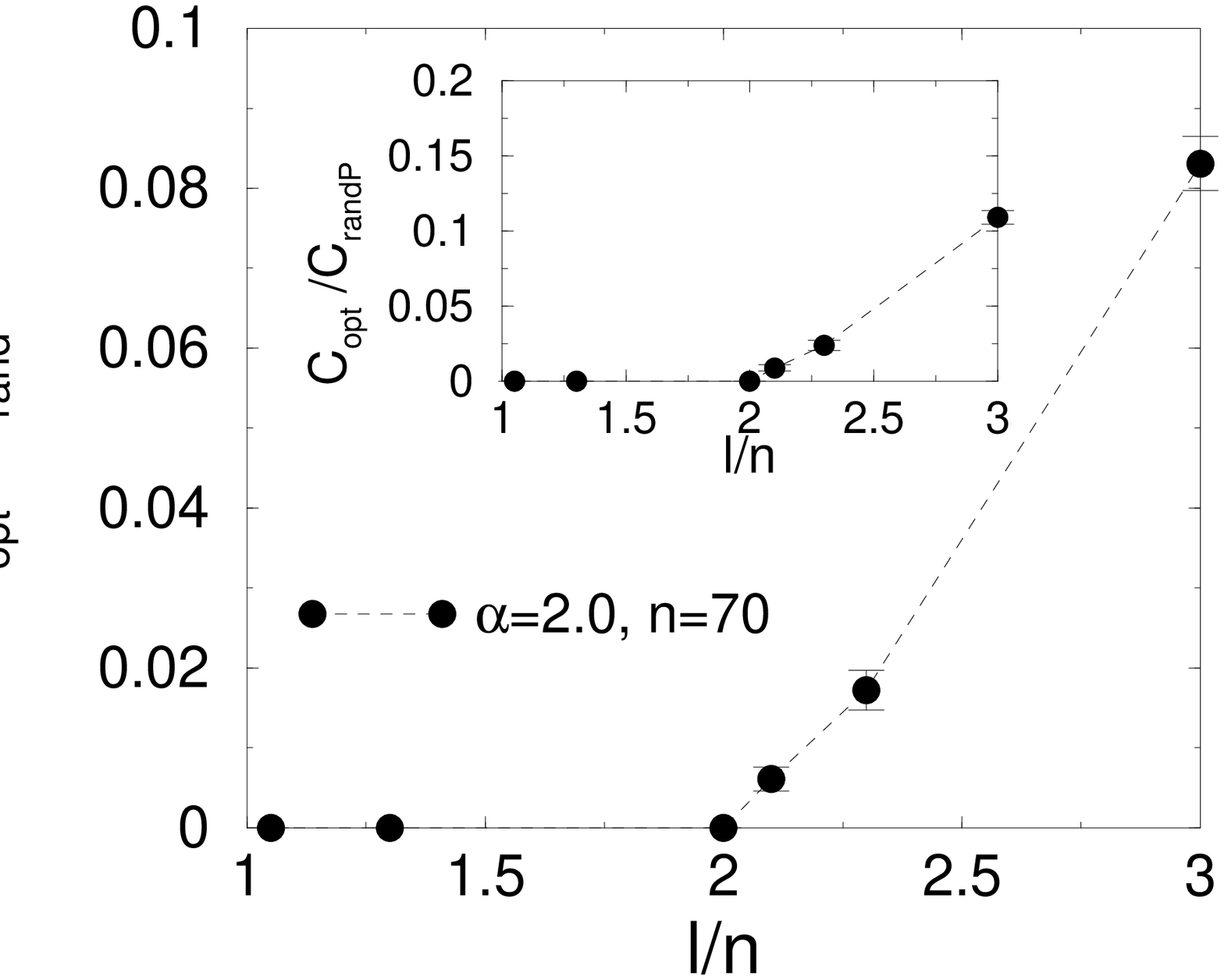}


    \includegraphics[width=5.5cm,height=4cm]{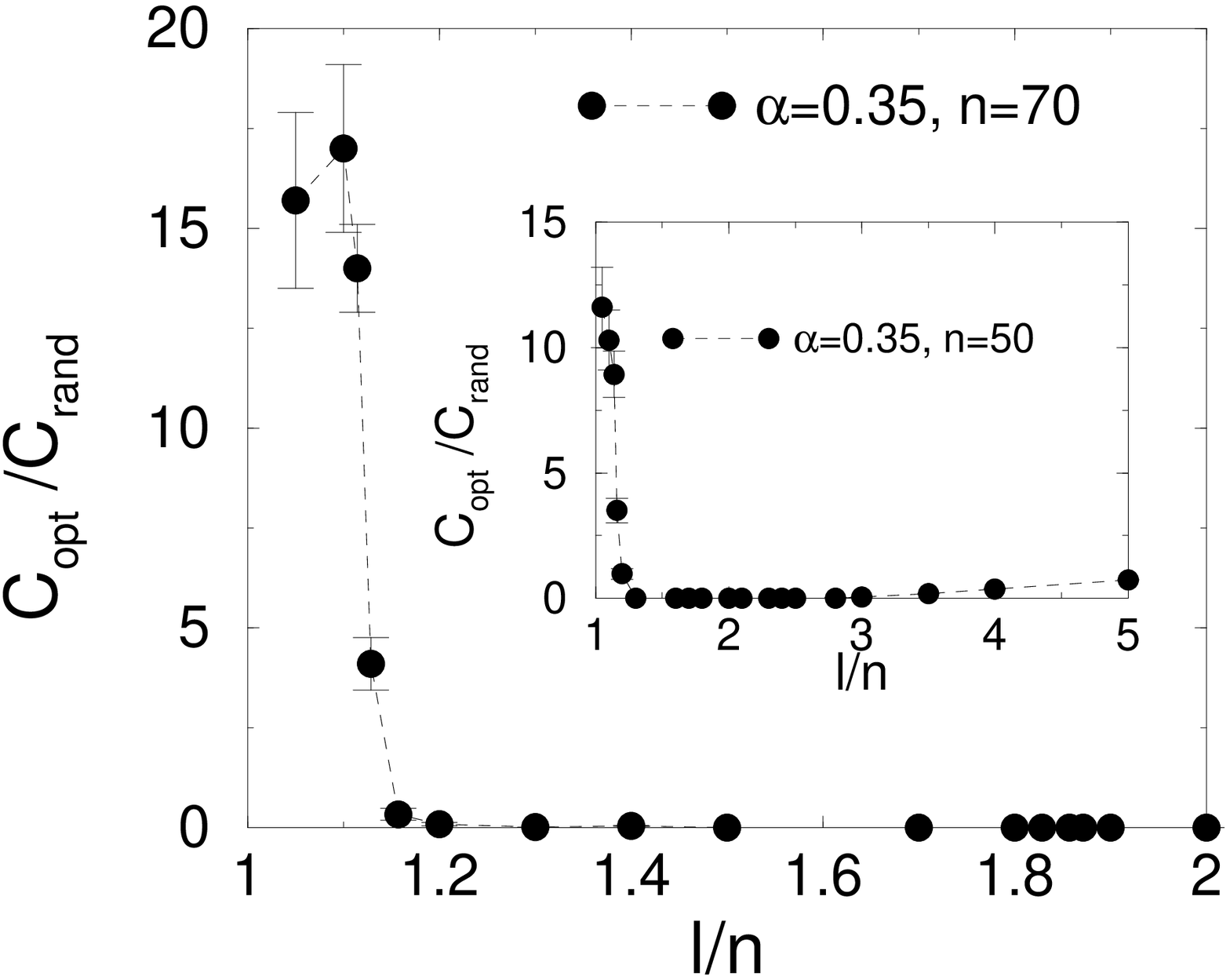}
\end{center}
\vspace{-.7cm} \caption{ Mean clustering coefficient 
for  the optimal configuration $C_{opt}$ normalized to 
the mean clustering coefficient, $C_{rand}$, of the random configuration. Top: results for network size 
$n=70$ and  $\alpha=2.0$; in the inset the behaviour of the ratio 
$C_{opt}/C_{randP}$ is shown, where $C_{randP}$ represents the mean clustering 
of a random graph with the same degree distribution $P(k)$ as the 
optimized  network. Bottom: results for network size 
$n=70$ and $\alpha=0.35$; in the inset ($n=50$, $\alpha=0.35$) both the 
critical values, $r_c(\alpha)$ and $r_c'(\alpha)$, are shown.}
\label{fig:C}
\end{figure}

For $\alpha>1$  (fig.~\ref{fig:C} Top), 
the system undergoes a clear phase transition as the value of the
ratio $r$ increases passing from a regime characterized by zero
clustering to one in which the clustering coefficient
becomes different from zero. 
The cost function in eq.(\ref{H}) has two competing forces: 
the minimization of the graph diameter and the
minimization of node degree. When $\alpha>1$  the
minimization of node degree dominates and the system
attempts to minimize the degree of each node resulting in
a peaked distribution around the mean value $<k>$,  with a 
non-trivial topology characterized by zero clustering and exhibiting
the presence of long loops. (fig.~\ref{fig:topologies} Bottom Right). 
When the ratio $r$
reaches the critical value $r_c(\alpha)$, one obtains
a non-zero clustering coefficient. 

 This transition also occurs for $\alpha<1$. 
However, when $\alpha<1$ one obtains an additional phase transition
at $r_c'(\alpha)$, where the system passes from optimal networks
exhibiting  a non-zero clustering coefficient, to ones with no
clustering at all. Starting from very small values of $r$, we
observe topologies characterized by the presence of  few interconnected
hubs (i.e. sites with very high degree \cite{complexnetworks,barab1})
linked to many peripheral sites (fig.~\ref{fig:topologies} Top Left).
Indeed, when $\alpha<1$, the tendency expressed by the cost function 
is to decrease the graph diameter, i.e. a measure of the
mutual distance among pairs of nodes. 

The emergence of this extra phase 
transition underscores the importance of the concavity (convexity) 
of the cost function.  

The limiting case $\alpha \to
0$ would correspond to the minimization of the standard graph distance,
leading,  in the region $r \sim 1$, to
a single central hub connected to  $n-1$
peripheral nodes, which share the
remaining $l-n+1$ links. This
situation  leads to non-zero clustering. 
The minimization of the  graph distance corresponds to a 
limiting case of \cite{sole} as well; however, in \cite{sole} there
is no constraint on the number of links $l$, so that the optimal network
they find  is a clique, in which each node is connected to
each other.

Increasing the ratio $r$ does not favour adding other links among
the hubs, because  their already high
degrees would only increase further. 
Hence the system reorganizes by increasing the number of
hubs and automatically reducing their degrees, trying to avoid
expensive triangles between hubs.  When the transition occurs, at 
$r_c'(\alpha)$, the
network does not exhibit hubs any more, but tends to become quite
homogeneous in the sense that almost every node has coordination
close to the average value $<k>$. Even in this regime the optimal
topology  is distinctly different from the random one. 
In fact, it displays
a peaked degree distribution around the mean value $<k>$ without
significant clustering (fig.~\ref{fig:topologies} Bottom Left).
The loops formed have the maximum possible length in order to reduce the
energy function. Adding extra links to the network forces  the
loops to become smaller, still avoiding clustering up to a
 second critical value of $r$, $r_c(\alpha)$. Beyond this value,
 'triangles' appear leading to  a transition similar 
to the one encountered for $\alpha>1$ 
(fig.~\ref{fig:C} Bottom, inset).

The extent of the clustering phase for $r<r_c'(\alpha)$ and $\alpha<1$ 
shrinks for increasing values of $\alpha$; the critical value $r_c(\alpha)$
decreases as $\alpha$ increases, $\forall \alpha$.
 From Fig.~\ref{fig:Pk_a0.7_1.05__a0.7_n70},~\ref{fig:Pk_n70_a0.7} 
and Fig.~\ref{fig:C}, one finds that several distinct  
topologies are obtained for different values of $\alpha$ and $r$:
a heterogeneous regime exhibiting a broad distribution of 
degrees ($r\sim 1$, $\alpha<1$) observable both in the
clustering and no clustering phase depending on the value of $\alpha$; 
a homogeneous regime for larger 
values of $r$ with $C \ne 0$ ($r>r_c(\alpha)\,\,\forall \alpha$, 
and $\alpha<1$,
$r<r_c'(\alpha)$ but not in the tree-like limit)
 or $C=0$ ($\alpha<1$, $r_c'(\alpha)<r<r_c(\alpha)$ and $\alpha>1$, 
$r<r_c(\alpha)$).

We have also
studied the characteristic path length, $L$, 
defined as the average, over all 
pairs in the system, of the graph distance between pairs of nodes.

\begin{figure}[!ht]
\begin{center}
    \includegraphics[width=5.5cm,height=4cm]{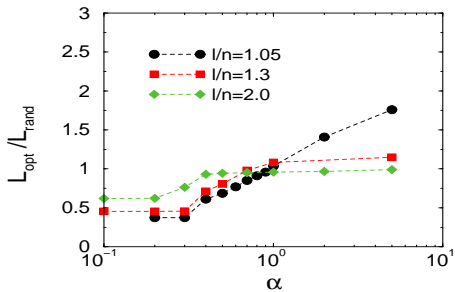}
\end{center}
\vspace{-.7cm} \caption{Characteristic path length $L_{opt}$,
normalized to the classical random one $L_{rand}$, vs. $\alpha$.}
\label{fig:L}
\end{figure}

 As shown in fig.~\ref{fig:L}, in the entire interval of $\alpha$,
the characteristic path length of the optimal
configuration, $L_{opt}$, is comparable to or smaller than
the random one, $L_{rand}$.  Even though the small network sizes studied 
here do not allow us to reach  definitive 
conclusions, the system  seems to  display a small-world effect
\cite{newman}.

 We have studied the system behaviour in terms of mean clustering and
average path length in comparison to both a classical random graph
\cite{erdos,bollobas} ($C_{opt}/C_{rand}$ and $L_{opt}/L_{rand}$) and a
random graph characterized by the same degree distribution $P(k)$ as the
optimized network ($C_{opt}/C_{randP}$ and $L_{opt}/L_{randP}$): both
studies give comparable results (see for example the top inset of
fig.~\ref{fig:C}).

In summary, we have investigated the role of selective pressure
in determining the topological
features observed in natural and artificial complex networks. Our 
work is complementary to
existing models that either rely on dynamical mechanisms, such as
preferential attachment, or on topological and geometrical
criteria.  Optimality leads to the emergence of
several distinct network structures
including an apparent scale-free arrangement in the tree-like topology limit.
Besides the degree distribution, we have studied the
clustering coefficient and the average path length of the selected
networks  which point to the existence of  non-trivial phase transitions
and to the features of the  small-world effect.
Our main result is  that the emergence of the topologies
observed in nature may not exclusively be the outcome of growth
mechanisms but may also arise from the interplay of dynamical
mechanisms with an evolutionary selection process.

This work was supported by COFIN MURST 2001, NASA
and by NSF IGERT grant DGE-998758.



\end{document}